\documentclass[apjl]{emulateapj}

\usepackage{natbib}
\newcommand{\chandra}{{\it Chandra}}

\newcommand{\s}[1]{\altaffilmark{#1}}

\shorttitle{Cosmic Rays in Tycho}
\shortauthors{Eriksen et al.}

\begin{document}

\submitted{Accepted to ApJ Letters}

\title{Evidence For Particle Acceleration to the Knee of the Cosmic Ray
Spectrum in Tycho's Supernova Remnant}

\author{Kristoffer A. Eriksen\s{1}, John P. Hughes\s{1}, Carles Badenes\s{2,3},
Robert Fesen\s{4}, Parviz Ghavamian\s{5}, David Moffett\s{6}, Paul P. Plucinksy\s{7},
Cara E. Rakowski\s{8}, Estela M. Reynoso\s{9,10},Patrick Slane\s{7}}

\altaffiltext{1}{Department of Physics and Astronomy, Rutgers, the State University of
New Jersey, Piscataway, NJ 08854, USA}
\altaffiltext{2}{School of Physics and Astronomy, Tel-Aviv University, Tel-Aviv 69978, Israel}
\altaffiltext{3}{Benoziyo Center for Astrophysics, Weizmann Institure of Science, Rehovot 76100, Israel}
\altaffiltext{4}{Department of Physics and Astronomy, Dartmouth College,Hanover, NH 03755, USA}
\altaffiltext{5}{Space Telescope Science Institute, Baltimore, MD 21218, USA}
\altaffiltext{6}{Department of Physics, Furman University, Greenville, SC 29613, USA}
\altaffiltext{7}{Harvard-Smithsonian Center for Astrophysics,Cambridge, MA 02138, USA}
\altaffiltext{8}{Space Science Division, Naval Research Laboratory,Washington, DC 20375, USA}
\altaffiltext{9}{Instituto de Astronomi\'a y Fi\'sica del Espacio,Buenos Aires, Argentina}
\altaffiltext{10}{Faculty of Exact and Natural Sciences, University of Buenos Aires, Argentina}

\begin{abstract}
Supernova remnants (SNRs) have long been assumed to be the source of
cosmic rays (CRs) up to the ``knee'' of the CR spectrum at 
$10^{15}\; \mathrm{eV}$, accelerating particles to relativistic energies
in their blast waves by the process of diffusive shock acceleration (DSA).
Since cosmic ray nuclei do not radiate efficiently, their presence
must be inferred indirectly. 
Previous theoretical calculations and X-ray observations show that 
CR acceleration modifies significantly the structure of the SNR and
greatly amplifies the interstellar magnetic field.
We present new, deep X-ray observations of the remnant
of Tycho's supernova (SN 1572, henceforth Tycho), which reveal a previously unknown,
strikingly ordered pattern
of non-thermal high-emissivity stripes in the projected interior of the remnant, with spacing that
corresponds to the gyroradii of $10^{14} - 10^{15}\; \mathrm{eV}$\ protons.
Spectroscopy of the stripes shows the plasma to be highly
turbulent on the (smaller) scale of the Larmor radii of TeV energy electrons.  
Models of the shock amplification of magnetic fields
produce structure on the scale of the gyroradius of the highest energy CRs 
present, but they do not predict the highly-ordered pattern
we observe. We interpret the stripes as evidence for acceleration of particles
to near the knee of the CR spectrum in regions of enhanced magnetic turbulence,
while the observed highly ordered pattern of these features provides a new challenge to models
of DSA.
\end{abstract}


\section{Introduction}
Energetic arguments have long-favored the hypothesis that supernova shocks are 
responsible for the bulk of the cosmic ray (CR) flux at Earth from 
$10^{8}\; \mathrm{eV}$\ 
to the ``knee'', a flattening of the CR spectrum
at approximately $10^{15}\; \mathrm{eV}$ \citep{BlandfordEichler1987}. 
This requires
that $\sim 10\%$ of the mechanical energy of Galactic supernova remnants (SNRs)
be converted into relativistic particles over the remnants' lifetime.
In the favored model of this process, diffusive shock acceleration (DSA),
self-generated magnetic turbulence
provides scattering centers from which suprathermal particles may diffuse
between the up- and downstream fluid, gaining energy (on average) with
each shock crossing \citep{Bell1978a}. 
The ubiquitous non-thermal radio emission in SNRs is evidence of GeV electrons in SNRs, 
and the X-ray synchrotron emission observed from the limbs of
young SNRs \citep{Koyama1995,Hughes2000_CasA,Gotthelf2001_CasA,Hwang2002_Tycho}
indicates the presence of electrons to tens of TeVs.
Though CR nuclei comprise the vast majority of mass in SNR relativistic
particles, they do not radiate efficiently, and their presence must be inferred indirectly,
either by the modification of the shock dynamics~\citep{Warren2005} and
thermal state of the plasma~\citep{Hughes2000_E0102,Decourchelle2000,PatnaudeEllisonSlane2009},
or by the production of secondary pions~\citep{Butt2001,Aharonian2006}. 

\begin{figure*}
\begin{center}
\includegraphics[scale=0.75]{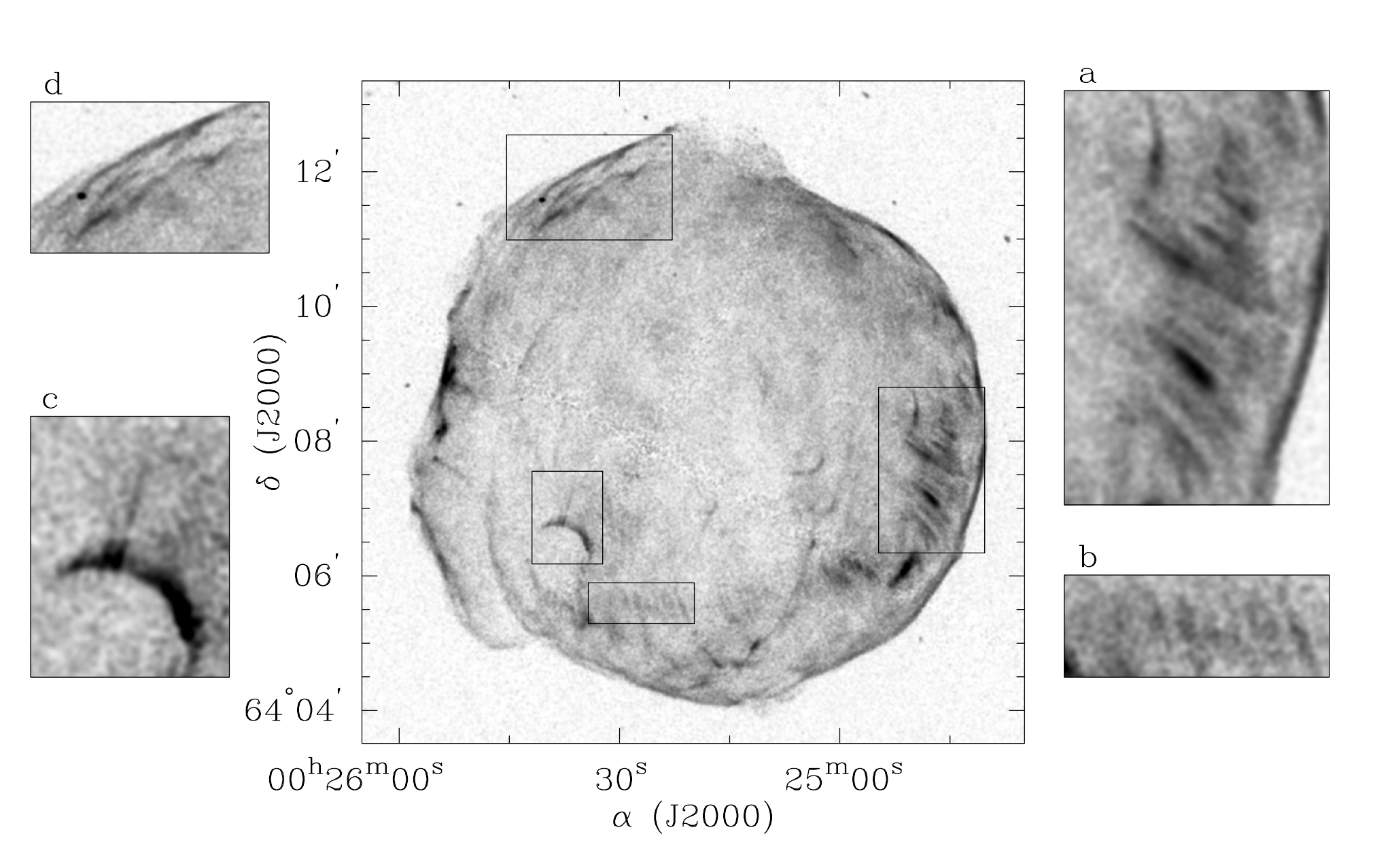}
\caption{\chandra\ X-ray 4.0--6.0 keV image of the Tycho supernova
remnant, smoothed with a $\sim 0.75$\arcsec\ Gaussian
and displayed with an {\it arcsinh} scaling,
showing various regions of striping in the
nonthermal emission. 
Clockwise from the upper right: a) The main western
stripes discussed in this Letter; b) A fainter ensemble of
stripes; c) a previously-known bright arc of non-thermal 
emission, with our newly discovered streamers; d)  filaments  of ``rippled sheet''
morphology common in optical observations of middle-aged SNRs. \label{ImageFig}}
\end{center}
\end{figure*}

\cite{BellLucek2001} predicted significant amplification 
of the magnetic field in supernova shocks by CR-induced turbulent processes, allowing
for acceleration to the knee within a SNR lifetime. This picture is supported by
high spatial resolution observations from the {\it Chandra X-ray Observatory}
that require high magnetic fields to explain the short synchrotron cooling time
implied by the geometrically thin rims of young 
remnants~\citep{VinkLaming2003,Ballet2006}.
An extensive theoretical literature has developed detailing a range of plasma instabilities that
might generate these high fields.
Of these, the cosmic ray current driven instability 
\citep[CRCD, often referred to as Bell's non-resonant mechanism,][]{Bell2004}, 
in which the magnetic field is amplified by
turbulence generated by CRs propagating ahead of the shock,
has received the most study, though other
processes (e.g., acoustic instabilities, resonant instabilities, etc.) have
been suggested.
MHD \citep{Bell2005} and particle-in-cell (PIC) simulations \citep{Riquelme2009} show
that the CRCD instability drives plasma motions which suppress growth on
small scales, until the dominant scale of the turbulence is comparable to
the Larmor radius of the highest energy cosmic rays. This results in a non-uniform
medium with structured, filamentary density and field enhancements
$\sim10$\ times greater than the mean quantities, surrounding cavities
of size approximately equal to the Larmor radius of the highest energy
particles present. Indeed, Bell's
analysis and simulations show that once the instability has saturated, the
medium consists of a network of holes evacuated of plasma and field,
enclosed by ``wandering filaments'' of high density and frozen-in field.  
One goal of observations is to discover evidence for structure in 
SNR shocks on this spatial scale.

\begin{figure*}[t]
\begin{center}
\includegraphics[scale=0.6]{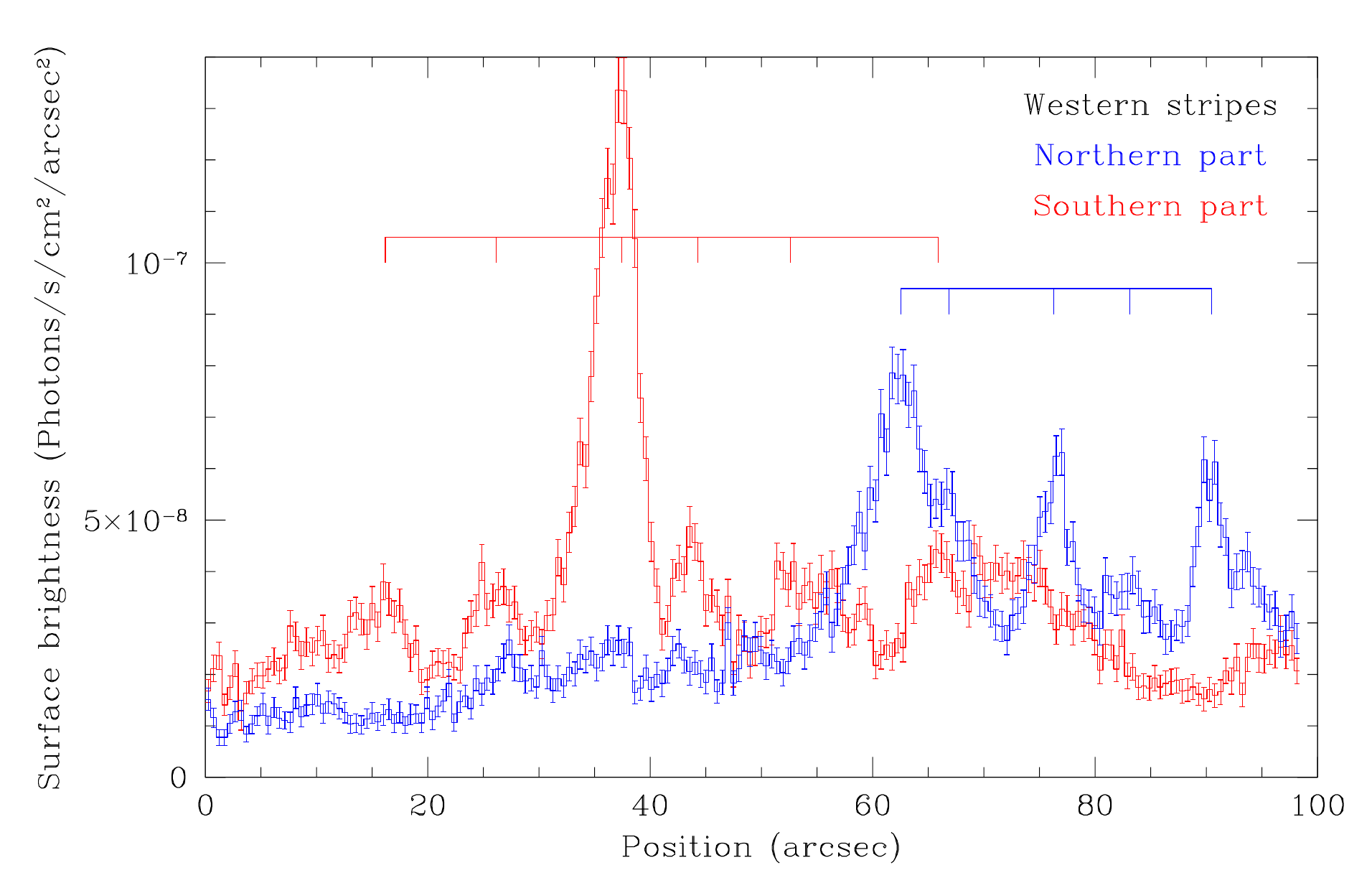}
\hbox{
\raisebox{0.35in}{ \includegraphics[scale=0.65]{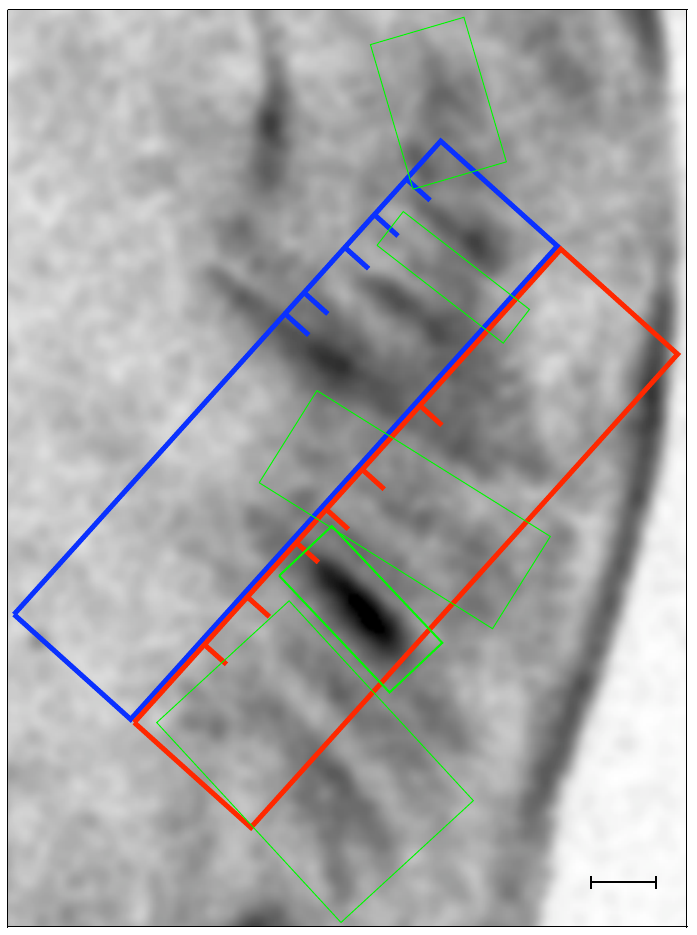} }
}
\caption{(Left) The projected intensity profile for two
 regions containing the western stripes.  The  bright stripe
 is the red peak around a position of 35$^{\prime\prime}$. The
 locations of prominent stripes are indicated by the horizontal lines
 with tick marks. The peaks in the northern profile are spaced by
 9.9$^{\prime\prime}$, 11.3$^{\prime\prime}$, 6.9$^{\prime\prime}$,
 8.3$^{\prime\prime}$, and 13.3$^{\prime\prime}$; those in the
 southern profile are spaced by 4.4$^{\prime\prime}$,
 9.4$^{\prime\prime}$, 6.8$^{\prime\prime}$, and 7.4$^{\prime\prime}$.
 (Right) Grayscale image of the \chandra\ data in a line-free region of
 the spectrum (4.2--6.0 keV) showing the extraction regions for the
 profiles plotted in the left panel.  The profiles run from the lower
 left to the upper right and the tick marks on the boxes correspond to
 the same tick marks as in the left panel. The scale bar in the lower
 right is 10$^{\prime\prime}$ long. The green boxes denote the spectral regions
analysed in the Letter. The brightest region is the bright stripe;
the coaddition of the fainter regions constitutes the faint stripes.
\label{StripeFig} }
\end{center}
\end{figure*}

\section{Observations}

We observed Tycho in April 2009 with the {\it Chandra X-ray Observatory}
Advanced CCD Imaging Spectrometer imaging array, 
as part of a Cycle 10 Large Program
(LP), using the 4 ACIS-I front-side illuminated CCDs, operated in Faint
mode.  The total program was split into nine individual obsids, which we
reprocessed with CIAO version 4.1, using the gain tables and CTI
correction in CALDB 4.1.3.  Examination of the lightcurves revealed no
significant background flares. The registration of the individual
pointings was improved using the measured relative positions of
background point sources, though the necessary shifts were typically
$\ll$1$^{\prime\prime}$.  For imaging analyses, the registered event
lists were merged using the standard CIAO tools, and have a total
average livetime of $\sim734.1\, {\rm ks}$. For spectroscopy,
counts were extracted and RMFs and ARFs were generated for each
individual obsid, which we fit jointly in XSPEC version 12.6.0.

In Figure \ref{ImageFig} we show the image from the 4--6 keV band,
which is dominated by the synchrotron component of the X-ray
spectrum. Apart from the well-known limb-brightened shell, a number of
bright regions are seen toward the projected interior of the
remnant. While the strongest of these features are visible in earlier
\chandra\ data~\citep{Warren2005} our deeper observation reveals a
striking pattern of nearly-regularly spaced stripes. The brightest
group, centered $\sim30$\arcsec\ interior to the western limb (see
Figure \ref{StripeFig}), has a peak surface brightness twice that of the brightest
sections of the rims, and is the primary subject of this
communication.  A second, fainter pattern extends east-west,
$55-75$\arcsec\ inside the southern rim, and there is evidence for
several other regions of striping near the detection limit.
Comparison with a shallower 2004 \chandra\ observation reveals no
statistically significant change in the brightness of the stripes,
ruling out any dramatic flux variability like that observed in the
non-thermal X-ray filaments of the SNR RX~J1713.7-3946~\citep{Uchiyama2007}.  
Our preliminary proper motion
measurements for the stripes are consistent with the overall expansion
of the blast wave and, in particular, show no evidence for non-radial
flow. There are no obvious counterparts to these features in the 
radio~\citep{Reynoso1997Tycho}, nor in the mid-IR.

\section{Analysis}

\subsection{Location of the Stripes}
Before investigating the nature of the stripes, we first must locate
them within the three-dimensional volume of the remnant.  The canonical
picture of a young SNR consists of three distinct fluid
discontinuities: the blast wave, which marks the shock propagating
into the ambient medium, a Rayleigh-Taylor (R-T) unstable contact
discontinuity (CD) at the ejecta-interstellar material boundary, and a
reverse shock that propagates into the stellar remains.  \cite{Warren2005} 
set an upper-limit for the azimuthally-dependent
projected radius of the reverse shock in Tycho using the location of
the Fe K$\alpha$\ emission.  Adopting their center of expansion, the
western stripes peak at a radius of 220\arcsec, well outside the
190\arcsec\ position of the reverse shock at that azimuth. While the
position of the stripes does coincide with the Warren et al.\ estimate
of the CD, the regularly-spaced, linear morphology of the non-thermal
stripes does not correspond to any features in the R-T plumes of thermal
emission tracing the ejecta boundary, nor is the CD a prominent feature 
elsewhere in the 4--6 keV band. Conversely, the blast wave is a bright source of
4--6 keV emission, and we identify the stripes as projected features
of this forward shock.

Tycho's blast wave is traced by a very thin shell of X-ray emission, with a
typical thickness only 1-2\% of its radius~\citep{Warren2005}. Since
the stripes are seen in projection away from the rim, their
line-of-sight pathlength through the shell is small. Thus their
intrinsic emissivity must be high relative to the limb regions.  We
examined two regions of different brightness on the western rim with
minimal ejecta contamination, and assumed a spherical shell
model with a radius of 214\arcsec, chosen to match the curvature of the western
limb.  The radial brightness profile of the brighter rim region is too
narrow and peaked to be consistent with a simple projected shell
geometry, meaning that its high surface brightness likely arises from
a local enhancement in emissivity, perhaps similar to the stripes. In
contrast, the profile of the fainter region is well-fit by a shell
with thickness 2\% of its radius, and is likely more typical of the
blast wave as a whole. This simple model predicts that features at
80\% of the shock radius (where the stripes are located) would have a
surface brightness 20\% that of the limb. The observed peak surface
brightness of the stripes is, however, 5 times that of the faint limb
region. Thus the peak intrinsic emissivity of the
synchrotron-emitting plasma in the stripes must be a factor of 25
higher than is typical for the blast wave. In principle, this
brightening could be due to a local enhancement in the ambient density and/or
magnetic field. 
While we can not entirely exclude this possibility, we consider the
existence of a pre-existing structure around Tycho
with the right combination of increased density, magnetic field, and, as we show
below, turbulence necessary to produce the observed correlations between
the ordered structure and spectral variations improbable.
In contrast, models of CR-driven magnetic field amplification
produce structure in the precursor density and field,
and we find it more likely that the stripes mark
a region extensively modified by the acceleration process.
This assumption is implicit in the analysis in this Letter.

\begin{figure*}[t!]
\begin{center}
\includegraphics[scale=0.625]{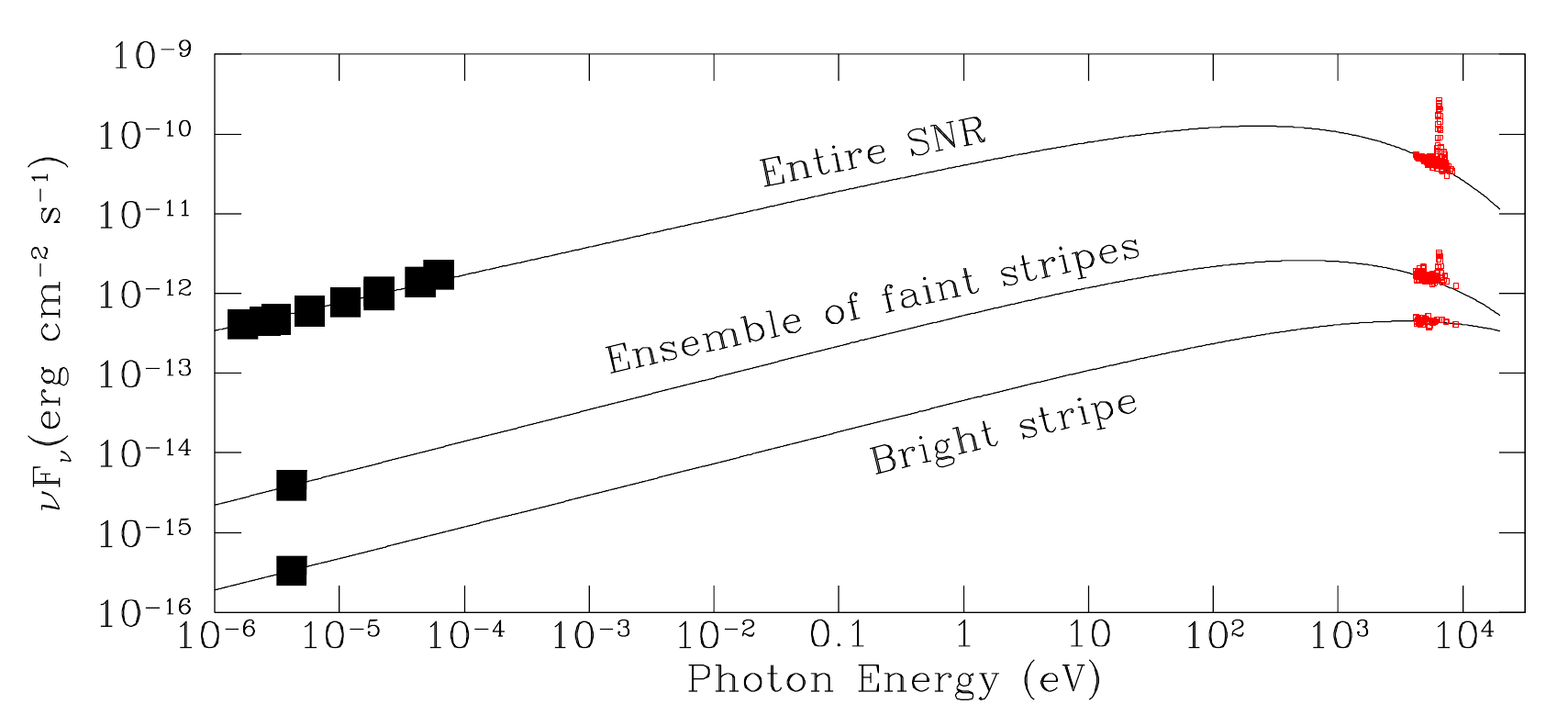}
\hbox{
\raisebox{0.05in}{ \includegraphics[scale=0.36]{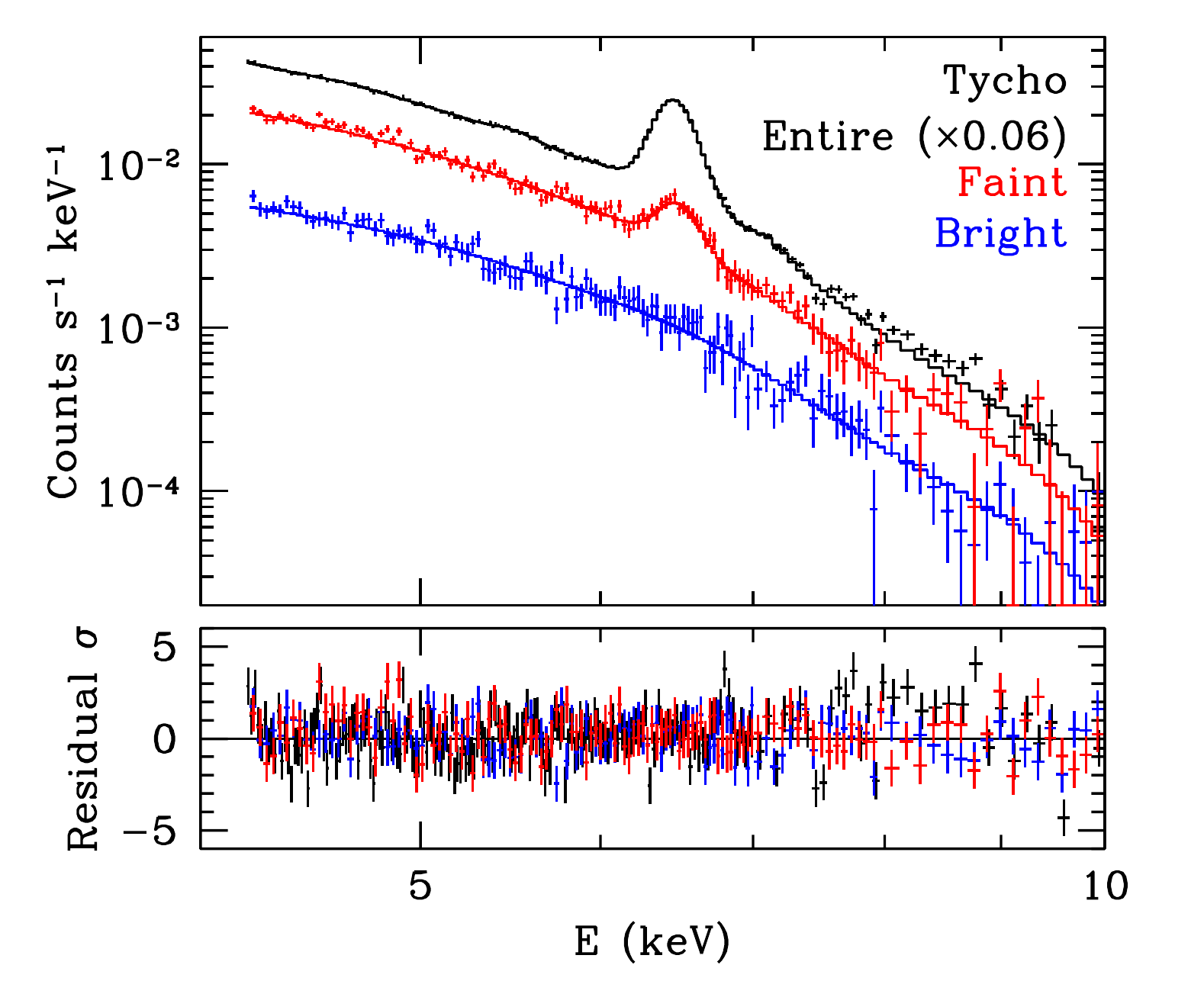} }
}
\caption{(Left) Broadband spectra of Tycho from the radio to
  the X-ray for the entire SNR (top), an ensemble of faint stripes
  (middle) and the bright stripe (bottom). The radio points for the
  latter two came from our own estimates based on the VLA radio data,
  taking account of the remnant's expansion.  The integrated radio
  flux density values for the entire SNR are taken from Klein et
  al.~(1979). The \chandra\ data are plotted in red and for the top two
  spectra the Fe K line is visible. The overplotted curves are the
  best-fit phenomenological model for the continuum emission of a
  power law (extrapolated from the radio band) with an exponential
  cut-off in the X-ray regime. (Right) \chandra\ spectra of the three
  spatial regions analyzed.  The top part shows the observed spectral
  intensity versus photon energy (data points with error bars) along
  with the best fit model. The bottom part shows the residuals (data
  $-$ model) in units of the flux error.  The data points and model
  curves here include instrumental effects, which have been removed for
  the X-ray data points plotted on the left panel.
\label{SpecFig}
}
\end{center}
\end{figure*}

\subsection{Spectroscopy}
The spatial position of the stripes, seen in projection against the
bright thermal X-rays of the shocked ejecta, greatly complicates
spectroscopy.
This can be mitigated by restricting analysis to the 4.2-10 keV range,
where the emission is
dominated by the synchrotron continuum and the Fe K$\alpha$\ line
complex at $\sim$6.45 keV. Throughout this work 
we set the absorbing column
to $N_H = 7 \times 10^{21}\; \mathrm{cm^{-2}}$, which has negligible
effect at these energies.
The spectrum of the brightest stripe is well-fit with a power-law of
photon index of $\Gamma = 2.11^{+0.08}_{-0.10}$, significantly harder
than the spectrum of the entire SNR at these energies 
($\Gamma = 2.72 \pm 0.02$; this fit includes faint lines identified by \citealp{Tamagawa2009}),
 and a coaddition of several of the lower
surface brightness stripes ($\Gamma = 2.52 \pm 0.05$; this fit
includes an Fe K line).

While photon indices are useful for characterization of the non-thermal spectrum,
the broadband spectrum gradually departs from a simple power law over several
decades in energy. Detailed DSA models of synchrotron spectra
show a small degree of curvature through the radio to UV,
with a cut-off in the X-rays, set (in young SNRs like Tycho) by the short cooling time
of TeV-scale electrons~\citep{EBB2000}.
Absent detailed spectral calculations for Tycho, we adopt a common 
approximation and assume a power-law electron momentum
distribution with an exponential cut-off, which produces a photon spectrum
that is nearly a power-law from the radio through the UV and is
curved in X-rays~\citep{ReynoldsSRCUT}.
For this model, 
the low energy index ($\alpha_{\mathrm{radio}}$) and the flux normalization
are set by radio observations, while the location of the high energy cut-off 
is constrained with the X-ray spectrum. 
For the spectrum of the entire remnant, assuming a radio spectral index of 
$\alpha_{\mathrm{radio}} = 0.65$, a flux density at 1 GHz of $\sim$50 Jy~\citep{Kothes2006},
and again including the faint X-ray emission lines 
from \cite{Tamagawa2009},  the fit to the \chandra\ data gives
$\nu_{\mathrm{cut}} = 1.9 \times 10^{17}\; \mathrm{Hz}\; (h\nu_{\mathrm{cut}}\; \sim0.79\, \mathrm{keV})$.
This provides a remarkably good fit ($\chi^2_{\nu} = 1.05$),
and is consistent with measurements from {\it Suzaku}.
(The statistical error  on this number is so small as to be irrelevant; 
the precision to which we measure it is
far greater than the efficacy of our emission model or the accuracy of 
the radio flux and spectral index measurements.)
If we instead assume $\alpha_{\mathrm{radio}} = 0.61$ \citep{Klein1979},
the fit is slightly worse ($\chi^{2}_{\nu} = 1.10$), 
and the cut-off moves to lower energies ($\nu_{\mathrm{cut}} = 0.9 \times 10^{17}\; \mathrm{Hz}\;; h\nu_{\mathrm{cut}}\,\; \sim0.37\, \mathrm{keV}$).
For the stripes, we estimated the radio flux in these regions from a VLA map.
The fits favor slightly shallower indices ($\alpha_{\mathrm{radio}} \sim 0.60$)
and higher cut-off energies,
$\nu_{\mathrm{cut}} = 19^{+13}_{-9}\times 10^{17}\; \mathrm{Hz}\; (7.9^{+5.4}_{-3.7}\, \mathrm{keV})$\ 
for the Bright Stripe and 
$\nu_{\mathrm{cut}} = 2.8 \pm 0.2 \times 10^{17}\; \mathrm{Hz}\; (1.16 \pm 0.04\, \mathrm{keV})$\
for the ensemble of fainter stripes.  
We plot the broadband synchrotron spectra of these regions in Figure \ref{SpecFig}.

Although it is clear that the Bright Stripe and Faint Ensemble spectra
require different best-fit parameter values, in principle these
differences might be due to either different cut-off
frequencies or radio spectral indices.  
Our spectral analysis (detailed in Table \ref{FitsTab})
allows us to discriminate between these two options.  
When the spectra for the Bright and Faint regions are fit with
radio flux densities fixed at their
nominal values, both regions prefer an index very close to the
average for the entire remnant. Conversely, if we instead require
that both regions have the same cut-off, 
the fitted radio indices are $\alpha = 0.565$\ (bright) and $\alpha =
0.623$\ (faint).  We cannot reject this level of variation outright since
measurements of the spatial variation of the radio spectral index are
not available for Tycho, though in Cas A (a plausible comparison
object) the radio index does vary spatially by much as $\pm
0.1$ \citep{Wright1999CasA}.  
On the other hand, the best fit for a single cut-off
frequency is clearly worse ($\Delta \chi^{2} = 13.8$) than for
independent cut-off values.  An $F$-test shows that the fit including
independent cut-off frequencies is an improvement at a significance
level of more than $3\sigma$.  Thus we conclude that the cut-off
frequencies are significantly different in the bright and faint
spectral regions analysed.

\section{Interpretation}
\subsection{The Diffusion Coefficient}
DSA requires magnetic field inhomogeneities strong enough that 
particles can scatter multiple times across the shock front.
This diffusion of CRs in the magnetized turbulence is 
generally assumed to be close to the Bohm regime,
where the mean-free path is the particle gyroradius.
Departure from this limit can be parameterized as $k_0 = D_0/D_{0,\mathrm{Bohm}}$,
where $D_0$\ is the diffusion coefficient at the electron cut-off energy.
\cite{Parizot2006} show that for an electron distribution limited by
synchrotron losses, the photon cut-off energy ($E_{\gamma,\, \mathrm{cut}} = h \nu_{\mathrm{cut}}$)
is set by the diffusion coefficient, shock velocity ($V_3$, in units of 1000 km s$^{-1}$), 
and compression ratio, $r$:
\begin{equation}
\label{k0}
k_0 = 0.14\; E^{-1}_{\gamma,\mathrm{cut,keV}}\; V^{2}_{3,\mathrm{shock}}\; \frac{16\,(r-1)}{3\,r^{2}}
\end{equation}
%
At a distance of $4.0 \pm 1.0\; \mathrm{kpc}$ \citep{Hayato2010}, with a blast wave proper motion of 
$0.3\arcsec\; \mathrm{yr^{-1}}$ \citep{Katsuda2010_Tycho}, $v_s = 5700\; \mathrm{km\, s^{-1}}$.
For an unmodified shock ($r = 4$) and our fit cut-off energies, 
$k_0 =$\ 5.8 (Whole SNR), $3.9 \pm 0.1$ (faint stripes), and 
$0.6^{+0.4}_{-0.3}$\ (Bright Stripe). 
With a higher compression more applicable to Tycho ($r = 7$), these numbers become
$3.8$, $2.6 \pm 0.1$, and $0.4^{+0.3}_{-0.2}$.
(The error bars reflect only the statistical error in $\nu_{\mathrm{cut}}$, and
not the $\sim 25\%$\ uncertainty in the distance to Tycho.)
Our value for the whole SNR is near to those found by
\cite{Parizot2006} for Tycho
($k_0 = 4.9-10$), while the very small numbers for the brightest
stripe are similar to their results for G347.3-0.5 and SN 1006
($k_0 = 0.2 - 0.9$). Since $k_0 > 1$\
for isotropic turbulence, this may be due to our oversimplified
emission model, and could indicate that the
photon spectrum falls more gradually to high energies, as seen in some 
DSA calculations \citep{Ellison2010}, or may be indicative of anisotropic
diffusion \citep{Reville2008}.
Regardless, our measured photon cut-off energies
require an order of magnitude decrease in the diffusion coefficient
in the stripes relative to the average value in Tycho, and  indicate that diffusion
in the stripes is very near the Bohm limit. Furthermore,
since Bohm diffusion requires $\delta{B}/B \sim 1$, $k_0 \sim 1$\ implies fully developed
turbulence, at least on the scale of the gyroradius of the highest energy ($\sim$ TeV)
electrons in the brightest stripes.

\subsection{The Origin of the Stripes}

Might the spatial structure and enhanced turbulence in the stripes be
a manifestation of the plasma instability driving the cosmic ray acceleration?
The largest possible characteristic scale of the acceleration process
is set by the gyroradius of the highest energy particles present.  If
we identify the gaps between the stripes ($l_{\rm gap}\sim 8\arcsec$) with twice
the proton gyroradius, an estimate of the magnetic field yields a
measurement of the energy, given by Equation~\ref{Ecr}.
\begin{equation}
\label{Ecr}
E_{\mathrm{CR}} = 9\;
\Bigl ( \frac{l_{\rm gap}}{1 \arcsec} \Bigr )
\Bigl ( \frac{D}{4.0\; \mathrm{kpc}} \Bigr )
\Bigl ( \frac{B}{\mathrm{\mu G}} \Bigr )
\times 10^{12}\; \mathrm{eV}
\end{equation}
\noindent
The choice of $B$\ is somewhat uncertain. If the stripes are a consequence
of the highest energy CRs interacting with
the non-amplified ISM field ($B \sim 3\; \mu\mathrm{G}$), for a gap spacing of
8\arcsec, $E_{CR} = 2 \times 10^{14}\; \mathrm{eV}$. However,
these features may arise from a region where the
field has already been somewhat amplified. 
For a shock velocity of $5700\; \mathrm{km\; s^{-1}}$, fits to models of Tycho's
synchrotron emission \citep{CC07} predict an
upstream field strength of $B \sim 30\; \mu \mathrm{G}$, indicating
$E_{CR} = 2 \times 10^{15}\; \mathrm{eV}$ --- just at the knee of the Galactic
CR spectrum.

\section{Conclusion}

If the Bell mechanism is active in the magnetic field amplification
in SNR shocks, and if the pattern of stripes we observe
in Tycho is an observable consequence of this process, the stripes are direct evidence
of particles accelerated to the knee. There are several caveats to consider
before accepting this conclusion. Most obviously, while MHD and PIC simulations 
of the CRCD
generate structure in the amplified medium, they do not show the regular and nearly
periodic pattern that we observe in Tycho. Secondly, when the CR back-reaction
is included \citep{Riquelme2009}, PIC simulations indicate that the Bell
instability alone may be neither fast nor efficient enough to provide
the observed amplification, particularly for Tycho's low ambient density.
Additional instabilities are required, and seem to be indicated
by the degree of turbulence in the stripes implied by our observations.
\cite{Riquelme2010} propose a new instability
(the Parallel Current Driven Instability, PCDI)
that acts on smaller scales (i.e., lower energies) and provides significant
field amplification over that established by the CRCD.
Further theoretical work on the degree of turbulence generated by
proposed mechanisms on TeV particle scales seems particularly warranted.
Regardless, the match in scales between the observed spacing of the Tycho
stripes and the Larmor radius of knee-region CRs --- a scale intrinsic to the 
CRCD --- is tantalizing, while the ordered structure 
presents a challenge to current
models of diffusive shock acceleration. 

\acknowledgments

This work was partially supported by \chandra\ award (grant number GO9-0078X)
to Rutgers University. 
EMR is member of the CIC (CONICET) and is funded by
the CONICET, UBA and ANPCyT projects.
JPH would like to acknowledge helpful discussions with Mario Riquelme, Anatoly
Spitkovsky, Tom Jones, Roger Blandford, and Martin Laming.  We also
thank Christopher McKee for contributing to the original \chandra\ 
 proposal.

{\it Facilities:} \facility{Chandra}.



\begin{deluxetable}{lrrrrr}
\tablecolumns{5}
\tablecaption{Stripe Spectroscopic Fits \label{FitsTab}}
\tablehead{
\colhead{Region} & \colhead{$\alpha$} & \colhead{$F_{\nu}$ (1 GHz)} & 
\colhead{$\nu_{\mathrm{cut}}$} & \colhead{$\chi^{2}/\mathrm{d.o.f.}$ }
\\
\colhead{} & \colhead{} & \colhead{(Jy)} & \colhead{(Hz)} & \colhead{} 
}
\startdata
Bright Stripe  & $0.61$* & $3.33\times 10^{-2}$* & $2.60\pm0.09\times 10^{18}$ & 194.89/205 \\
Bright Stripe  & $0.603^{+0.010}_{-0.025}$ &  $3.33\times 10^{-2}$*  & $1.88^{+1.25}_{-0.89}\times 10^{18}$ & 194.30/204 \\
Bright Stripe  & $0.637^{+0.014}_{-0.012}$ &  $6.66\times 10^{-2}$* &  $1.93^{+2.09}_{-0.74}\times 10^{18}$ & 194.26/204 \\
Bright Stripe  & $0.65$* &  $8.0^{+2.3}_{-1.5}\times 10^{-2}$ &  $2.37^{+1.64}_{-1.00}\times 10^{18}$ & 194.27/204 \\
Bright Stripe  & $0.61$* &  $4.2^{+0.8}_{-1.1}\times 10^{-2}$ & $1.56^{+1.53}_{-0.44}\times 10^{18}$  & 194.27/204 \\
Faint Ensemble & $0.61$* & $0.39$*                   & $3.52\pm0.03\times 10^{17}$ & 667.29/687 \\
Faint Ensemble & $0.601^{+0.009}_{-0.004}$ &  $0.39$* & $2.84\pm0.10\times 10^{17}$ & 666.31/686 \\
Faint Ensemble & $0.635^{+0.011}_{-0.004}$ &  $0.78$* & $2.88\pm0.11\times 10^{17}$ & 666.55/686 \\
Faint Ensemble & $0.65$* &  $0.899^{+0.032}_{-0.026}$  &  $3.53\pm0.23\times 10^{17}$ & 666.37/686 \\
Faint Ensemble & $0.61$* &  $0.394\pm0.012$         &  $3.47^{+0.22}_{-0.36}\times 10^{17}$  & 667.25/686 
\enddata
\tablenotetext{*}{Fixed parameter}
\end{deluxetable}

\end{document}